\documentclass[conference]{IEEEtran}
\IEEEoverridecommandlockouts
\usepackage{cite}
\usepackage{amsmath,amssymb,amsfonts}
\usepackage{algorithmic}
\usepackage{graphicx}
\usepackage{textcomp}
\usepackage{xcolor}
\usepackage{url}
\def\BibTeX{{\rm B\kern-.05em{\sc i\kern-.025em b}\kern-.08em
    T\kern-.1667em\lower.7ex\hbox{E}\kern-.125emX}}

\usepackage{booktabs}
\usepackage{caption}
\usepackage[table]{xcolor}
\usepackage{multirow}
\usepackage{makecell}

\usepackage{tikz}
\newcommand\copyrightnotice{%
\begin{tikzpicture}[remember picture,overlay]
\node[anchor=south, yshift=30pt] at (current page.south) [fill=white, inner sep=2pt] {%
  \parbox{\textwidth}{\footnotesize Presented at the 2025 IEEE 22nd India Council International Conference (INDICON). \copyright 2025 IEEE. Personal use of this material is permitted. For all other uses, permission must be obtained from IEEE.}};
\end{tikzpicture}%
}

\begin{document}

\title{AUDRON: A Deep Learning Framework with Fused Acoustic Signatures for Drone Type Recognition
\thanks{AmygdalaAI-India Lab, is an international volunteer-run research group that advocates for \textit{AI for a better tomorrow} https://amygdalaaiindia.github.io/.}
}

\author{\IEEEauthorblockN{Rajdeep~Chatterjee$^{1,3}$,Sudip~Chakrabarty$^{1,3}$*, Trishaani~Acharjee$^{1,3}$, Deepanjali~Mishra$^{2}$
}

\IEEEauthorblockA{$^1$School of Computer Engineering, KIIT Deemed to be University, Bhubaneswar 751024, India\\
$^2$School of Liberal Studies, KIIT Deemed to be University, Bhubaneswar 751024, India\\
$^3$Amygdala AI, Bhubaneswar 751024, India\\
Email: \{cse.rajdeep, sudipchakrabarty6, trishaaniacharjee8\}@gmail.com }} 

\thispagestyle{plain}
\makeatother
\maketitle
\copyrightnotice

\begin{abstract}
Unmanned aerial vehicles (UAVs), commonly known as drones, are increasingly used across diverse domains, including logistics, agriculture, surveillance, and defense. While these systems provide numerous benefits, their misuse raises safety and security concerns, making effective detection mechanisms essential. Acoustic sensing offers a low-cost and non-intrusive alternative to vision or radar-based detection, as drone propellers generate distinctive sound patterns. This study introduces AUDRON (AUdio-based Drone Recognition Network), a hybrid deep learning framework for drone sound detection, employing a combination of Mel-Frequency Cepstral Coefficients (MFCC), Short-Time Fourier Transform (STFT) spectrograms processed with convolutional neural networks (CNNs), recurrent layers for temporal modeling, and autoencoder-based representations. Feature-level fusion integrates complementary information before classification. Experimental evaluation demonstrates that AUDRON effectively differentiates drone acoustic signatures from background noise, achieving high accuracy while maintaining generalizability across varying conditions. AUDRON achieves 98.51\% and 97.11\% accuracy in binary and multiclass classification. The results highlight the advantage of combining multiple feature representations with deep learning for reliable acoustic drone detection, suggesting the framework’s potential for deployment in security and surveillance applications where visual or radar sensing may be limited.
\end{abstract}


\section{Introduction}
The rapid advancement of drone technology has transformed multiple industries, enabling innovative applications in logistics, precision agriculture, infrastructure inspection, environmental monitoring, and security \cite{tahir2025future}. Modern UAVs are increasingly compact, agile, and affordable, which allows for flexible deployment but also increases the risk of misuse \cite{jackman2025everyday}. Unauthorized drones can intrude into restricted airspaces, perform covert surveillance, or disrupt public events, creating significant safety, privacy, and regulatory challenges. Efficient detection and identification of such UAVs in real time is therefore crucial for operational safety, security enforcement, and public protection.

Traditional detection techniques \cite{chen2025review}, including radar, LiDAR, and vision-based systems, offer high accuracy under controlled conditions but often face limitations in practical scenarios. Small or fast-moving drones can evade radar, while poor lighting, occlusions, or cluttered environments reduce the effectiveness of camera-based systems. Moreover, these approaches can be costly and infrastructure-intensive, making them less feasible for widespread deployment. Acoustic sensing emerges as a low-cost and non-intrusive alternative, leveraging the distinct sound signatures produced by drone propellers. However, environmental noise, overlapping sound sources, and variations in drone models pose challenges to reliable acoustic detection, necessitating robust feature extraction and intelligent modeling techniques.

This study introduces AUDRON, which stands for AUdio-based Drone Recognition Network, as a hybrid deep learning framework to address these challenges. AUDRON integrates multiple feature representations, including MFCC, STFT spectrograms processed with CNN, recurrent layers for temporal sequence modeling, and autoencoder-based embeddings for capturing latent audio characteristics. Feature-level fusion combines these complementary representations, enhancing the system’s capability to discriminate drone acoustic signatures from diverse background noises.

Key contributions include:

\begin{itemize}
    \item Proposal of a unified framework (AUDRON) that effectively integrates multiple complementary feature representations for drone acoustic detection.
    \item Demonstration of model robustness using diverse drone and environmental sounds, incorporating a noise class to minimize false positives in real-world scenarios.
\end{itemize}

The paper is organized as follows. Section II reviews related work on drone classification. Section III details the proposed AUDRON model. Section IV presents the experimental setup, results, and discussion. Section V concludes with future directions.

\section{Related Works}
Several studies have explored methods for capturing and analyzing drone sounds from real-world environments. While numerous studies focus on object detection using drones and drone imagery \cite{10969335}, comparatively fewer works address drone detection specifically from the distinctive acoustic signatures they generate. Al-Emadi et al. \cite{al2019audio} proposed a deep learning approach using acoustic fingerprints to detect and identify drones from recorded audio. Alla et al. \cite{alla2024sound} proposed an audio-visual fusion approach combining CRNN and YOLOv5 for drone detection, achieving high accuracy, though reliance on both modalities may limit performance in scenarios where either audio or IR data is severely degraded. Zhang et al. \cite{7966780} developed an audio-assisted camera array that fused visual and acoustic features for drone detection, though its dependence on bulky multi-camera setups limits portability and scalability. Iqbal et al. \cite{8616877} proposed a sound-based amateur drone detection framework using MFCC \cite{11173395} and Linear Predictive Cepstral Coefficients (LPCC) features \cite{article1} with SVM classifiers, achieving high accuracy but showing sensitivity to noisy environments. Dong et al. \cite{dong2023drone} proposed a deep learning-based drone sound detection system using fused acoustic features. Akbal et al. \cite{AKBAL2023104012} proposed a sound-based amateur drone detection model using Skinny Pattern and Iterative Neighborhood Component Analysis(INCA) feature selection, achieving high accuracy on a small multi-class environmental sound dataset. This work \cite{11007784} developed a deep learning-based UAV audio detection and identification system using STFT spectrograms, evaluating CNN, RNN, and CRNN models on diverse drone and environmental sounds for robust real-world performance. This work uses a dataset that includes both drone sounds and diverse environmental acoustic sounds as a no-drone class, aligning with environmental sound classification \cite{10829485} approaches to better handle real-world false positives. These studies highlight the growing role of acoustic cues in drone detection, yet also underscore the need for more diverse datasets and robust methods to address real-world variability and false alarms.

\section{Proposed Methodology}
\subsection{Dataset Description}
\subsubsection{Synthetic Drone Audio Generation}

\begin{figure}[ht!]
    \centering
    \begin{minipage}[]{0.48\textwidth}
        \centering
        \includegraphics[width=\textwidth, height=6cm]{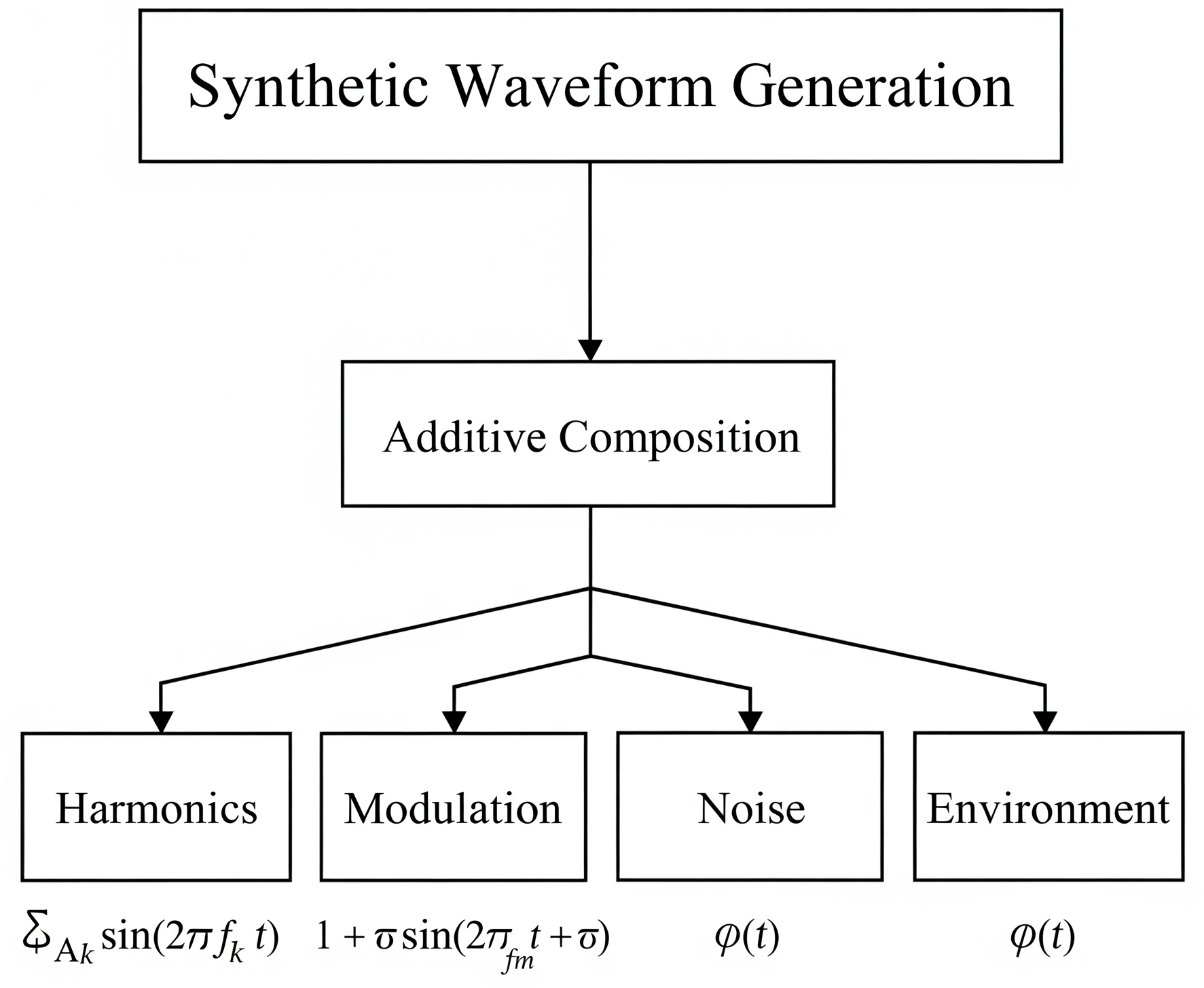}
        \caption{Flowchart of synthetic audio generation highlighting key components used in the process.}
        \vspace{-4mm}
        \label{fig:synth_flowchart}
    \end{minipage}\hfill
\end{figure}

To augment the training data and improve model robustness, synthetic drone audio signals were generated. The process, illustrated in Figure~\ref{fig:synth_flowchart}, models a realistic waveform by combining the core harmonic signature of the drone's rotors with modulation, noise, and environmental factors.

For a given drone class, such as a quadcopter, hexacopter, octocopter, or racing\_drone, the audio waveform $x_c(t)$ of duration $T$ is modeled as follows:
\begin{equation}
x_c(t) = \left( \sum_{k=1}^{K_c} A_k \sin(2 \pi f_k t) \right) \cdot M(t) + \eta(t) + \xi(t), \quad 0 \le t \le T
\label{eq:synth_drone_audio}
\end{equation}
where:
\begin{itemize}
    \item $A_k$ and $f_k$ denote the amplitude and frequency of the $k^{th}$ harmonic of the drone’s rotors. These are varied per class to simulate distinct rotor configurations:
    \begin{itemize}
        \item \textbf{Quadcopter}: 4 rotors, base frequency $\approx$ 75 Hz
        \item \textbf{Hexacopter}: 6 rotors, base frequency $\approx$ 65 Hz
        \item \textbf{Octocopter}: 8 rotors, base frequency $\approx$ 55 Hz
        \item \textbf{Racing drone}: Base frequency $\approx$ 120 Hz
    \end{itemize}
    \item $K_c$ is the number of harmonics considered for class $c$.
    \item $M(t) = 1 + \alpha \sin(2 \pi f_m t + \phi)$ represents low-frequency RPM modulation, simulating throttle variations (where $\alpha \approx 0.15$ and $f_m \approx 1.5$ Hz).
    \item $\eta(t)$ is additive Gaussian noise, modeling background disturbances.
    \item $\xi(t)$ captures low-frequency wind or other environmental effects.
\end{itemize}

\begin{figure*}[ht]
    \centering
    \begin{minipage}[t]{0.99\textwidth}
        \centering
        \includegraphics[width=16.5cm, height=5cm]{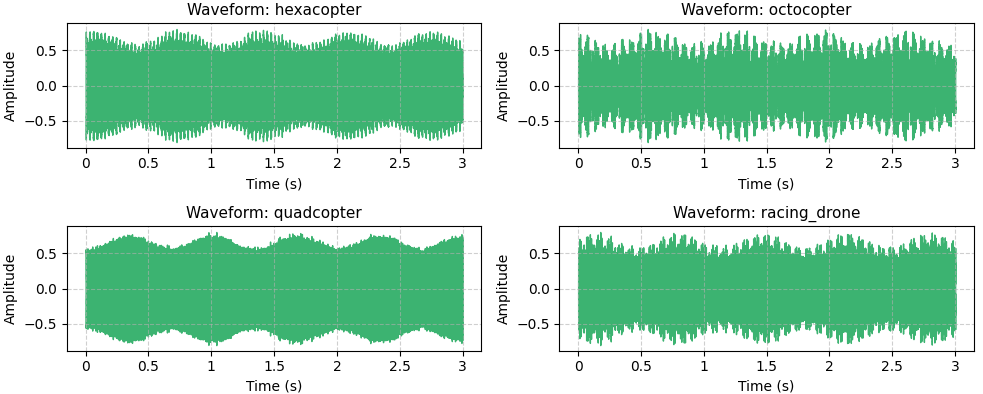}
        \caption{Sample waveforms from the synthetic model, illustrating distinct audio signatures for the four drone classes.}
        \label{fig:custom_data}
    \end{minipage}\hfill

\vspace{2mm}
    \centering
    \begin{minipage}[t]{0.99\textwidth}
        \centering
        \includegraphics[width=\textwidth, height=11cm]{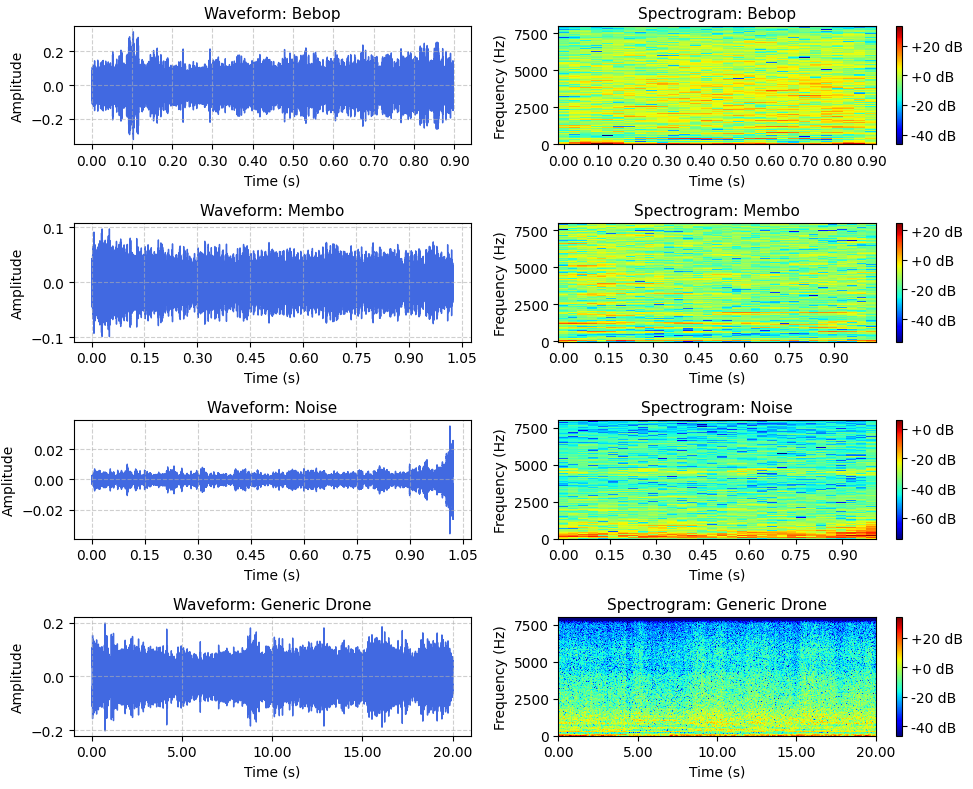}
        \caption{Sample waveforms (left) and spectrograms (right) from the experimental datasets.}
        \label{fig:sample_data}
    \end{minipage}\hfill
\end{figure*}

This approach generates realistic waveforms for each drone type, enabling extraction of features like MFCCs and STFT spectrograms for training our detection framework (Figure~\ref{fig:custom_data}).

\subsubsection{Real-World Datasets}
The study uses the DroneAudioDataset \cite{al2019audio} of indoor drone propeller recordings. To enhance robustness, it is augmented with environmental noise from ESC-50 \cite{piczak2015esc}, white noise from Speech Commands, and silence clips to reduce bias. Sample waveforms and spectrograms of drone and noise recordings are shown in Figure~\ref{fig:sample_data}.

To further strengthen the model's performance for binary classification (drone vs. non-drone), we specifically enlarged the drone class with a supplementary dataset \cite{RamosRomero2023} containing 175 unique drone audio files. This augmentation step was crucial for exposing the model to a wider variety of drone acoustic signatures, thereby improving its generalization and reducing the risk of overfitting. Sample waveforms and spectrograms of the generic drone data are illustrated in Figure~\ref{fig:sample_data}.

\subsection{AUDRON Model Architecture}
\begin{figure*}[ht]
    \centering
    \begin{minipage}[t]{0.99\textwidth}
        \centering
        \includegraphics[width=\textwidth, height=10.5cm]{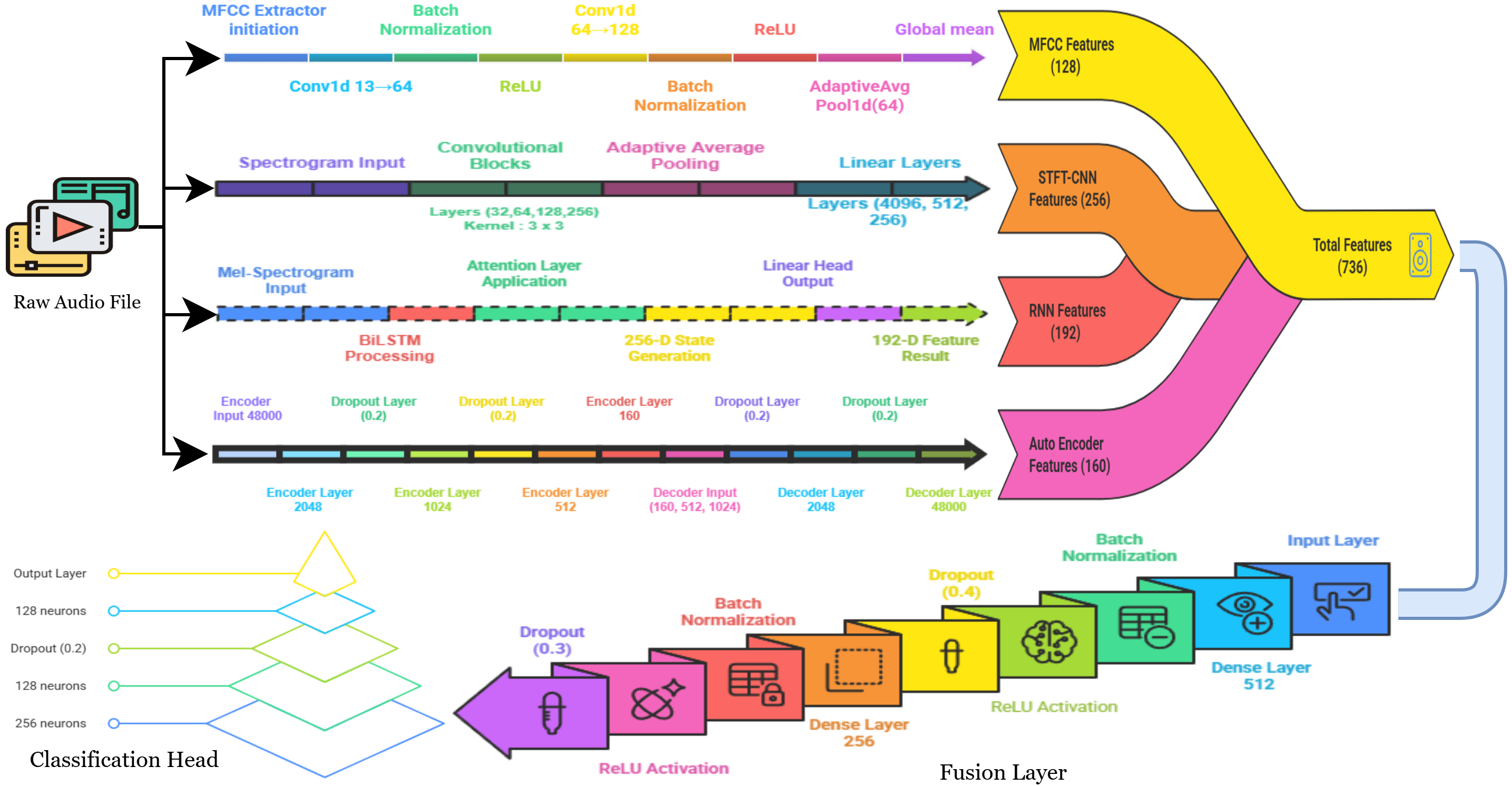}
        \caption{The complete architecture of our proposed multi-modal drone detection model, AUDRON. A raw audio waveform is processed by four parallel feature extraction branches: MFCC extractor, STFT-CNN, RNN, and Autoencoder. The resulting feature vectors are concatenated and passed through a fusion layer and final classification head to produce the output.}
        \label{fig:model_architecture}
    \end{minipage}\hfill
\end{figure*}

The proposed AUDRON model is a multi-modal architecture designed to robustly classify drone audio. As illustrated in Figure~\ref{fig:model_architecture}, the model processes a raw audio waveform through four parallel branches to extract a diverse set of features, which are then synergistically combined for a final prediction.

\begin{itemize}
    \item \textbf{MFCC Extractor:} This branch captures the timbral characteristics of the audio using MFCCs and a 1D Convolutional Neural Network (CNN). It specifically calculates 13 MFCCs and processes them through two sequential `Conv1d' layers to produce a 128-dimensional feature vector representing spectral texture.

    \item \textbf{STFT-CNN Extractor:} This branch treats the audio like an image by generating an STFT spectrogram and feeding it to a deep 2D CNN. Four Conv2d blocks learn hierarchical patterns, effectively capturing complex spectro-temporal signatures.

    \item \textbf{RNN Extractor:} Focusing on temporal dependencies, this branch uses a bidirectional LSTM with attention. It captures context from past and future audio segments while emphasizing the most informative time steps.

    \item \textbf{Audio Autoencoder:} This branch learns a compressed representation of the raw audio in a self-supervised manner. The encoder reduces the 48,000-sample input to a 160-dimensional embedding, capturing essential features.

    \item \textbf{Fusion and Classification Head:} The feature vectors from all four branches are concatenated into a 736-dimensional vector and passed through a dense fusion layer with Dropout and Batch Normalization, before the final classification head.
\end{itemize}

\begin{figure*}[ht]
    \centering
    \begin{minipage}[t]{0.99\textwidth}
        \centering
        \includegraphics[width=17cm, height=13cm]{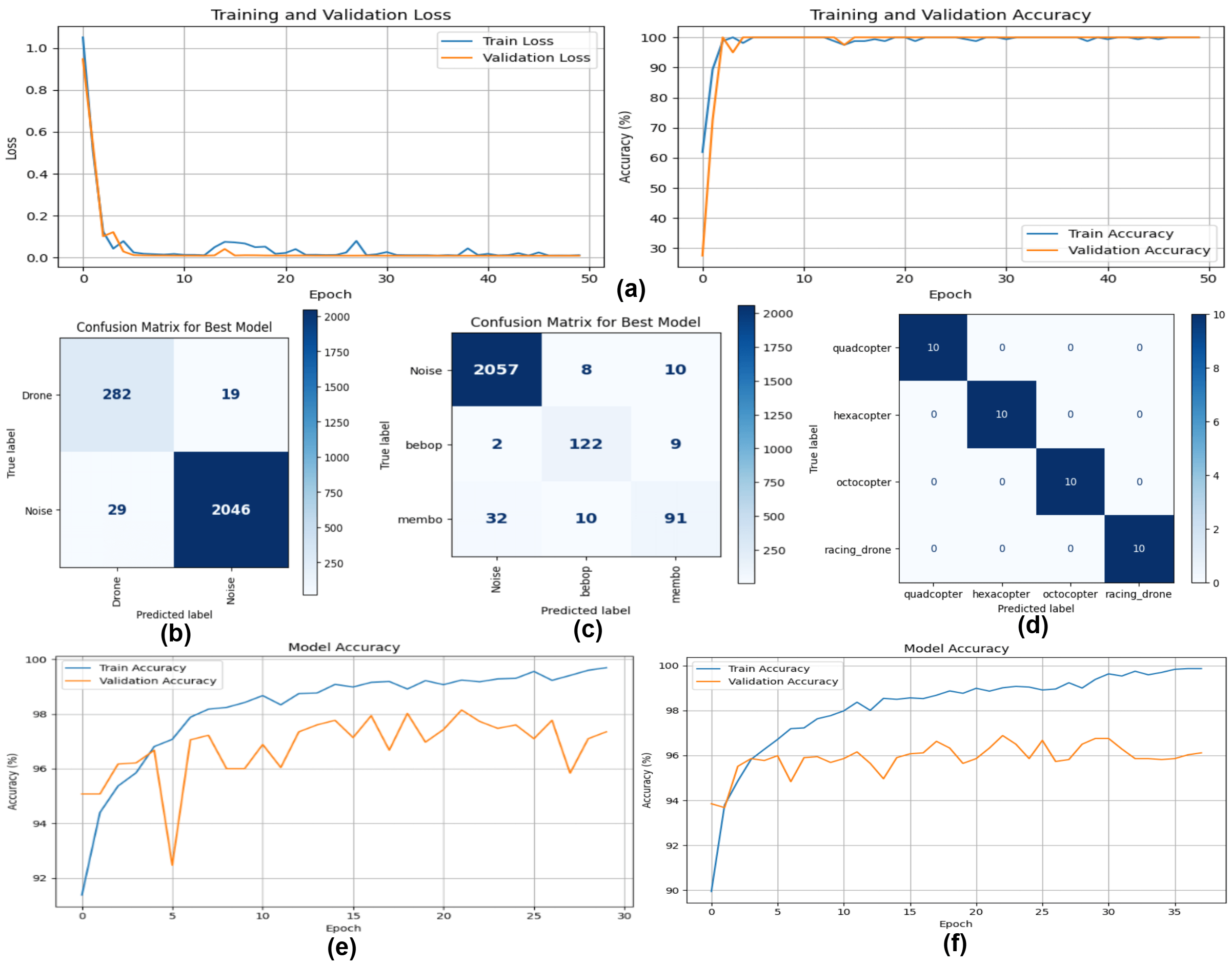}
        \caption{Training dynamics and performance evaluation of the proposed model across different datasets and tasks. (a) Training and validation loss/accuracy curves for the model trained on synthetically generated data, showing rapid convergence. (b) Confusion matrix for binary classification on real-world data. (c) Confusion matrix for the multiclass classification task. (d) Confusion matrix for the model evaluated on the synthetic dataset. (e) Training and validation accuracy curves for the binary classification model. (f) Training and validation accuracy curves for the multiclass classification model.}
        \label{fig:model_output}
    \end{minipage}\hfill
\end{figure*}

\subsection{Training Setup}
All models were trained and evaluated on the Kaggle platform using an NVIDIA T4 GPU. Training ran for up to 50 epochs with a batch size of 16. We used the AdamW optimizer with an initial learning rate of 0.001 and a ReduceLROnPlateau scheduler to decrease the learning rate if validation accuracy did not improve for 5 consecutive epochs. The loss function combined Cross-Entropy Loss for classification and a weighted Mean Squared Error (MSE) for the autoencoder reconstruction. Early stopping was used to prevent overfitting, and the model with the highest validation accuracy was saved for final evaluation. Data splits for training and validation are shown in Table~\ref{tab:data_split_full}.

\begin{table}[h!]
\caption{Detailed distribution of audio samples for each experimental setup. Abbreviations are: B (Bebop), D (Drone), M (Membo), N (Noise).}
\label{tab:data_split_full}
\begin{tabular}{llcl}
\toprule
\textbf{Task} & \textbf{Split} & \textbf{Samples} & \textbf{Per-Class Distribution} \\ 
\midrule
\multirow{2}{*}{\makecell[l]{Binary \\ (No Augmentation)}} & Training & 9,363 & 1,066 D, 8,297 N \\
 & Validation & 2,341 & 266 D, 2,075 N \\
\midrule
\multirow{2}{*}{\makecell[l]{Binary \\ (with Augmentation)}} & Training & 9,502 & 1,205 D, 8,297 N \\
 & Validation & 2,376 & 301 D, 2,075 N \\
\midrule
\multirow{2}{*}{Multiclass} & Training & 9,363 & 533 B, 533 M, 8,297 N \\
 & Validation & 2,341 & 133 B, 133 M, 2,075 N \\
\midrule
\multirow{2}{*}{\makecell[l]{Multiclass \\ (Synthetic Data)}} & Training & 260 & -- \\
 & Validation & 80 & -- \\
\bottomrule
\end{tabular}
\end{table}

\subsection{Evaluation Metrics}
The performance of all models was evaluated using standard classification metrics. The reported metrics include overall Accuracy for general correctness, Precision to measure the reliability of positive predictions, and Recall to assess the model's ability to identify all positive instances. To account for potential class imbalance in the dataset, the F1-Score is also utilized, as it provides a balanced assessment by combining both Precision and Recall.

\section{Results and Discussion}
\subsection{Overall Model Performance}
The performance of the AUDRON model across various experimental configurations is shown in Table~\ref{tab:performance_comparison}. On Synthetic Data, it achieved near-perfect results with 99.92\% accuracy, validating its capability on clean generated signals. For the real-world binary classification task, the highest performance was without augmentation, reaching 98.51\% accuracy and an F1-Score of 0.9837. Adding augmented data slightly reduced accuracy to 97.98\%. In the multiclass task, the model maintained strong performance with 97.11\% accuracy and 0.9686 F1-Score. Training times increased with data complexity. To maintain the core objective of evaluating the AUDRON architecture, detailed noise-specific analysis was kept outside the scope of this study and is considered a direction for future work.

\subsection{Performance Visualization}
The training dynamics and classification performance of AUDRON are shown in Figure~\ref{fig:model_output}. Training and validation curves show rapid convergence, with high accuracy reached within the first few epochs across setups (a, e, f). Validation closely follows training, indicating good regularization and minimal overfitting. Confusion matrices provide detailed insights. For binary classification (b), 282 drone instances were correctly identified, with few noise misclassifications. High precision is also seen in multiclass tasks on real (c) and synthetic (d) data, where dominant diagonal entries indicate low class confusion.

\begin{table*}[h]
\centering
\caption{Model performance across different datasets and classification tasks.}
\label{tab:performance_comparison}
\begin{tabular}{lccccc}
\toprule
\cellcolor{blue!20}\textbf{Task / Data Configuration} & 
\cellcolor{green!20}\textbf{Accuracy (\%)} & 
\cellcolor{orange!20}\textbf{Precision} & 
\cellcolor{cyan!20}\textbf{Recall} & 
\cellcolor{magenta!20}\textbf{F1-Score} &
\cellcolor{red!20}\textbf{Training Time (min)} \\
\midrule
Synthetic Data & 99.92 $\pm$ 0.05 & 0.9990 $\pm$ 0.08 & 0.9995 $\pm$ 0.04 & 0.9989 $\pm$ 0.08 & 74 \\
\midrule

Binary Classification (No Augmentation) & \cellcolor{gray!20}\textbf{98.51 $\pm$ 0.09} & \cellcolor{gray!20}\textbf{0.9842 $\pm$ 0.11} & \cellcolor{gray!20}\textbf{0.9840 $\pm$ 0.10} & \cellcolor{gray!20}\textbf{0.9837 $\pm$ 0.12} & 235 \\
\midrule

Binary Classification (With Augmentation) & 97.98 $\pm$ 0.18 & 0.9802 $\pm$ 0.20 & 0.9799 $\pm$ 0.19 & 0.9799 $\pm$ 0.21 & 310 \\
\midrule

Multiclass Classification & 97.11 $\pm$ 0.25 & 0.9689 $\pm$ 0.28 & 0.9697 $\pm$ 0.26 & 0.9686 $\pm$ 0.07 & 293 \\
\bottomrule
\end{tabular}
\end{table*}

\begin{table*}[h!]
\centering
\caption{Ablation study results showing model performance with different feature combinations.}
\label{tab:ablation_combinations}
\begin{tabular}{lccccc}
\toprule
\cellcolor{blue!20}\textbf{Model Configuration} & 
\cellcolor{green!20}\textbf{Accuracy (\%)} & 
\cellcolor{orange!20}\textbf{Precision} & 
\cellcolor{cyan!20}\textbf{Recall} & 
\cellcolor{magenta!20}\textbf{F1-Score} &
\cellcolor{red!20}\textbf{Perf. Drop (\%)} \\
\midrule

STFT-CNN + RNN + Autoencoder & 93.81 $\pm$ 0.31 & 0.9321 $\pm$ 0.24 & 0.9334 $\pm$ 0.25 & 0.9333 $\pm$ 0.22 & \textbf{4.70} \\
MFCC + RNN + Autoencoder & 97.38 $\pm$ 0.19 & 0.9740 $\pm$ 0.21 & 0.9714 $\pm$ 0.18 & 0.9352 $\pm$ 0.20 & \textbf{1.13} \\
MFCC + STFT-CNN + Autoencoder & 97.61 $\pm$ 0.15 & 0.9755 $\pm$ 0.19 & 0.9761 $\pm$ 0.17 & 0.9757 $\pm$ 0.16 & \textbf{0.90} \\
MFCC + STFT-CNN + RNN & 97.92 $\pm$ 0.12 & 0.9787 $\pm$ 0.16 & 0.9791 $\pm$ 0.14 & 0.9792 $\pm$ 0.13 & \textbf{0.59} \\
\midrule
\textbf{MFCC+STFT+RNN+Autoencoder} & \cellcolor{gray!20}\textbf{98.51 $\pm$ 0.09} & \cellcolor{gray!20}\textbf{0.9842 $\pm$ 0.11} & \cellcolor{gray!20}\textbf{0.9840 $\pm$ 0.10} & \cellcolor{gray!20}\textbf{0.9837 $\pm$ 0.12} & \textbf{---} \\
\bottomrule
\end{tabular}
\end{table*}

\subsection{Ablation Study}
The ablation study (Table~\ref{tab:ablation_combinations}) validated the contribution of each AUDRON feature branch. All four branches improved performance, confirming the multi-modal fusion approach. The MFCC branch was most critical, with its removal causing a 4.70\% accuracy drop, highlighting its spectral texture importance. RNN and STFT-CNN removals reduced accuracy by 0.90\% and 1.13\%, respectively, while the Autoencoder added 0.59\%. Combining all four branches achieved the highest accuracy of 98.51\%. This study highlights how combining multiple feature branches improves drone detection performance.

\subsection{Comparison with State-of-the-Art Methods}
As shown in the state-of-the-art comparison in Table~\ref{tab:model_comparison}, our proposed AUDRON model consistently and significantly outperforms the baseline CNN, RNN, and CRNN architectures across both binary and multiclass classification tasks. In the Binary Classification task, AUDRON achieves a top-tier accuracy of 98.51\%, surpassing the next best model (CNN) by over 2\%.  For the more challenging Multiclass Classification task, AUDRON again demonstrates its superiority, achieving an accuracy of 97.11\%, which is a substantial improvement of over 4\% compared to the best-performing baseline (CNN). These results show that AUDRON’s multi-modal architecture effectively captures drone audio features for accurate and robust detection.

\begin{table}[h!]
\centering
\caption{Performance comparison of the proposed model against baseline architectures without data augmentation.}
\label{tab:model_comparison}
\begin{tabular}{llcc}
\toprule
\cellcolor{blue!20}\textbf{Classification Task} & 
\cellcolor{red!20}\textbf{Model} & 
\cellcolor{green!20}\textbf{Accuracy (\%)} & 
\cellcolor{magenta!20}\textbf{F1-Score (\%)} \\
\midrule

\multirow{4}{*}{\textbf{Binary}} & CNN \cite{al2019audio} & 96.38 $\pm$ 0.69 & 95.90 $\pm$ 0.78 \\
& RNN \cite{al2019audio} & 75.00 $\pm$ 6.60 & 68.38 $\pm$ 8.16 \\
& CRNN \cite{al2019audio} & 94.72 $\pm$ 1.36 & 93.93 $\pm$ 1.61 \\
\cmidrule{2-4}
& \textbf{AUDRON} & \cellcolor{gray!20}\textbf{98.51 $\pm$ 0.09} & \cellcolor{gray!20}\textbf{98.37 $\pm$ 12.0} \\
\midrule
\multirow{4}{*}{\textbf{Multiclass }} & CNN \cite{al2019audio} & 92.94 $\pm$ 11.89 & 92.63 $\pm$ 1.320 \\
& RNN \cite{al2019audio} & 57.16 $\pm$ 11.33 & 55.62 $\pm$ 13.53 \\
& CRNN \cite{al2019audio} & 92.22 $\pm$ 1.030 & 92.25 $\pm$ 1.010 \\
\cmidrule{2-4}
& \textbf{AUDRON} & \cellcolor{gray!20}\textbf{97.11 $\pm$ 0.251} & \cellcolor{gray!20}\textbf{96.71 $\pm$ 7.013} \\
\bottomrule
\end{tabular}
\end{table}

\section{Conclusion}
This research introduces AUDRON, a novel multi-modal deep learning architecture for acoustic drone detection. A single audio waveform is processed through four parallel feature extraction branches (MFCC, STFT-CNN, RNN, Autoencoder) to capture a comprehensive set of spectral, temporal, and latent features. Fusion of these representations enables the model to outperform baseline CNN, RNN, and CRNN architectures, achieving 98.51\% accuracy in binary detection. Ablation studies confirm that the combined feature streams drive its superior performance. Future work includes evaluating AUDRON in noisy real-world environments, expanding the dataset with diverse drone types, and developing a real-time, low-power version for deployment on edge devices, enabling practical acoustic surveillance of sensitive airspace.

\bibliographystyle{IEEEtran}
\bibliography{reference}

\end{document}